\documentclass[aps,prl,twocolumn,showpacs,superscriptaddress,groupedaddress]{revtex4}  
\usepackage{graphicx}  
\usepackage{dcolumn}   
\usepackage{bm}        
\usepackage{amssymb}   
\usepackage{amsmath}
\begin{document}

\title{Cycloidally modulated magnetic order stabilized by thermal fluctuations in the N\'{e}el-type skyrmion host GaV$_4$S$_8$}

\author{J.S. White}
\affiliation{Laboratory for Neutron Scattering and Imaging (LNS), Paul Scherrer Institut (PSI), CH-5232 Villigen, Switzerland}
\author{\'A. Butykai}
\affiliation{Department of Physics, Budapest University of Technology and Economics and MTA-BME Lend\"ulet Magneto-optical Spectroscopy Research Group, 1111 Budapest, Hungary}
\author{R. Cubitt}
\affiliation{Institut Laue-Langevin, 71 avenue des Martyrs, CS 20156, 38042 Grenoble cedex 9, France}
\author{D. Honecker}
\affiliation{Institut Laue-Langevin, 71 avenue des Martyrs, CS 20156, 38042 Grenoble cedex 9, France}
\author{C.D. Dewhurst}
\affiliation{Institut Laue-Langevin, 71 avenue des Martyrs, CS 20156, 38042 Grenoble cedex 9, France}
\author{L. F. Kiss}
\affiliation{Department of Experimental Solid State Physics, Institute for Solid State Physics and Optics, Wigner-MTA Research Centre for Physics, 1121 Budapest, Hungary}
\author{V. Tsurkan}
\affiliation{Experimental Physics V, Center for Electronic Correlations and Magnetism, University of Augsburg, 86135 Augsburg, Germany}
\affiliation{Institute of Applied Physics, Academy of Sciences of Moldova, Academiei str. 5, Chisinau, R. Moldova}
\author{S. Bord\'acs}
\affiliation{Department of Physics, Budapest University of Technology and Economics and MTA-BME Lend\"ulet Magneto-optical Spectroscopy Research Group, 1111 Budapest, Hungary}

\date{\today}

\begin{abstract}

We report small-angle neutron scattering studies of the lacunar spinel GaV$_4$S$_8$, which reveal the long-wavelength magnetic states to be cycloidally modulated. This provides direct support for the formation of N\'eel-type skyrmions recently claimed to exist in this compound. In striking contrast with all other bulk skyrmion host materials, upon cooling the modulated magnetic states transform into a ferromagnetic state. These results indicate all of the modulated states in GaV$_4$S$_8$, including the skyrmion state, gain their stability from thermal fluctuations, while at lower temperature the ferromagnetic state emerges in accord with the strong easy-axis magnetic anisotropy. In the vicinity of the transition between the ferromagnetic and modulated states, both a phase coexistence and a soliton-like state are also evidenced by our study.
\end{abstract}


\maketitle
Particle-like magnetic skyrmions with topologically non-trivial spin textures receive continued attention since their creation, stability and annihilation, either individually or in skyrmion lattices (SkLs), raise fundamental questions related to topology in physics \cite{Milde2013,Romming2013,Koshibae2014}. Moreover, the observations of metastable skyrmions \cite{Yu2010,Karube2016} and their current-driven dynamics \cite{Jonietz2010,Yu2012} inspires new ideas for using skyrmions in data storage and logic devices \cite{Nagaosa2013,Sampaio2013,Zhan2015}.

The spin texture of a skyrmion can be described by a vector field $\mathbf{S}$=S(cos($\Phi$($\varphi$))sin($\Theta$($\rho$)), sin($\Phi$($\varphi$))sin($\Theta$($\rho$)), cos($\Theta$($\rho$)), where $\varphi$ and $\rho$ are the azimuth and radial spatial coordinates, respectively \cite{Nagaosa2013,Bogdanov}. The non-trivial skyrmion topology is evidenced when its spin pattern is mapped on to a sphere, and the mapping wraps the surface of the sphere entirely. Two parameters classify the skyrmion type: the vorticity $m$ which is a non-zero integer, and the helicity, $\gamma$ where $\Phi(\varphi)=m\varphi+\gamma$.

The bulk cubic chiral helimagnets with $B$20 structure, like MnSi and FeGe, host whirlpool-like Bloch-type skyrmions described by $m$=+1 and $\gamma$=$\pm$$\pi$/2 \cite{Yu2010,Bogdanov,Muhlbauer2009}. In these compounds the competition between a dominant ferromagnetic (FM) exchange and weaker Dzyaloshinskii-Moriya interactions (DMIs) stabilize a long-wavelength ($\sim$10-100\,nm) helical ground state in zero magnetic field ($B$). The SkL phase exists under a finite $B$, where it competes with either a spin-flop conical phase or a field polarized state. Since the SkL is stable typically only over the narrow temperature ($T$) range $0.9\leq T\leq T_{C}$, where $T_{C}$ is the ordering $T$, thermal fluctuations are proposed to be crucial for the phase stability \cite{Bogdanov,Muhlbauer2009,Janoschek2013}.

Recently a new bulk SkL phase was reported in the lacunar spinel GaV$_4$S$_8$ (GVS) \cite{Kezsmarki2015}. At high $T$, GVS has a rock-salt like structure composed of alternating GaS$_4$ and V$_4$S$_4$ clusters. The highest-energy unpaired electron of the V$_4$S$_4$ cluster occupies a triply degenerate molecular orbital which drives a cubic ($F\overline{4}3m$) to rhombohedral ($R3m$) Jahn-Teller transition at $T_{JT}$=42\,K \cite{Pocha2000,Ruff2015,Wang2015,Widmann2016,Hlinka2016}. Since this phase transition breaks the $\overline{4}$ symmetry \cite{Pocha2000}, four structural domains develop below $T_{JT}$. Each domain carries an electric polarization along its rhombohedral axis, which lies parallel to one of the four cubic $\langle$111$\rangle$ directions \cite{Ruff2015,Widmann2016,Butykai2017}. Below $T_{JT}$, the Curie-Weiss $T$ indicates predominantly FM interactions between $S=$~1/2 spins of V$_4$S$_4$ clusters, and the system displays a strong easy-axis magnetic anisotropy~\cite{Kezsmarki2015,Widmann2016}. In contrast to a uniform FM state however, earlier small-angle neutron scattering (SANS) studies revealed that below $T_{C}$=13\,K, long-wavelength spiral order develops with a periodicity of 17\,nm at $T$=11\,K~\cite{Kezsmarki2015}.

Due to the rhombohedral $C_{3v}$ point symmetry of GVS at low $T$, the DMI pattern is expected to support a N\'eel-type SkL state with $m$=+1 and $\gamma$=0 or $\pi$~\cite{Bogdanov,Kezsmarki2015}, instead of a Bloch-type SkL. Correspondingly, the zero-field spiral state of GVS should be cycloidally and not helically modulated. However, microscopic experimental evidence for cycloidal order in GVS remains missing up to now. In addition, the SkL phase in GVS is reported to extend over a broader $T$ range down to 0.68\,$T_{C}$, while surprisingly, the spiral order at zero field is stable only \emph{above} $T_S$$\sim$5\,K, below which a FM state prevails. Hence, the key open questions concern the detailed nature of both the spiral order and phase stability in GVS, with this system attracting special interest as a hitherto unique bulk host of N\'eel-type skyrmions.

In this Letter we report comprehensive SANS studies of the long-wavelength magnetic order in GVS. We show that the zero-field spiral state is indeed cycloidally-modulated, underlining that in a finite field the N\'eel-type SkL is formed. In striking contrast to the $B$20 materials where a helical order forms the ground state as $T\rightarrow~0$, in GVS we observe the cycloidal order to collapse as the system becomes FM below $T_{S}=$~5\,K. We thus conclude GVS to display a delicate interplay between thermal fluctuations which stabilise $all$ of the modulated states closer to $T_{C}$, and a strong easy-axis magnetic anisotropy that drives the formation of the FM state at low $T$.



\begin{figure}[t]
\includegraphics[width=0.46\textwidth]{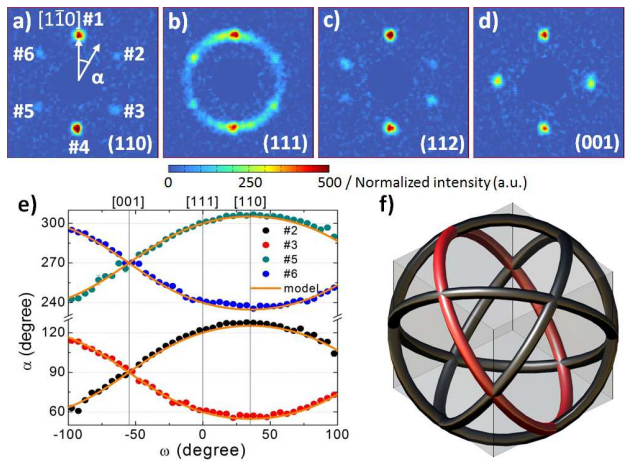}
\caption{(color online) Typical SANS data from GVS at 11\,K and in 0\,T, and recorded in the a) (110), b) (111), c) (112) and d) (001) planes, respectively. e) Angular position ($\alpha$ as defined in panel a)) of intensity spots \#2, \#3, \#5 and \#6 as a function of rocking angle $\omega$ around the vertical [1$\overline{1}$0] axis. f) Schematic image of the deduced intensity distribution in reciprocal space. An intensity ring confined to a \{111\} plane appears in each rhombohedral domain, which describes well the positions of the spots (orange lines) in panel e). The ring for the [111] domain is displayed in red only for clarity.}
\label{Fig1}
\end{figure}


Our SANS experiments were done using a 25\,mg single crystal of GVS, with the details of the crystal growth reported in Ref.~\cite{Kezsmarki2015}. Unpolarized SANS experiments were performed using the SANS-I instrument of the Paul Scherrer Institut and at D33 instrument of the Institut Laue-Langevin. Polarized SANS was also implemented at D33 as described in Ref.~\cite{Dewhurst2008}. Full details of our experimental are given in the Supplemental Material \cite{Supplement}. 

First, unpolarised SANS was used to map the distribution of scattered intensity from the spiral phase of GVS at $T$=11\,K and $B$=0\,T. SANS data were collected over a broad range of angles as the crystal was rotated (`rocked') around its vertical [1$\overline{1}$0] axis. Along with the two spots aligned with [1$\overline{1}$0] and [$\overline{1}$10]---which are on the rotation axis and so appear at all rocking angles---four other spots are also observed at all rocking angles with their relative angular separation changing smoothly [Figs.~\ref{Fig1}(a)-(e)]. As shown in Figs.~\ref{Fig1}(b) and (e), in the (111) image plane the six spots form a regular hexagon, while upon rotating towards (001) the angle between spots \#2-\#5 and \#3-\#6 decreases and they eventually merge in the (001) plane to form a four-spot pattern [Fig.~1(d)].

Besides the six spots, a ring of intensity is also observed lying in the (111) and ($\overline{1}$$\overline{1}$1) planes - see Fig.~\ref{Fig1}(b) for the (111) plane. The angular positions of the spot pairs \#2-\#5 and \#3-\#6 can also be ascribed to two additional intensity rings in the ($\overline{1}$1$\overline{1}$) and (1$\overline{1}$$\overline{1}$) planes that appear as spots when intersecting the Ewald sphere at the SANS detector. The rocking angle-dependence of the spot positions are consistently described by this model (see orange lines in Fig.~\ref{Fig1}(e)). Overall our analysis reveals that in GVS the scattered intensity is not concentrated into Bragg spots, but instead it is distributed over four homogeneous rings. Each ring corresponds to magnetic scattering from one of the four structural domains, with the scattering vectors $\textbf{q}$ confined to the rhombohedral planes, as shown in Fig.~\ref{Fig1}(f).

The assignment of intensity rings allows the selective analysis of the magnetic correlations within the different structural domains, even while the orientations of $\textbf{q}$ are ill-defined within the rhombohedral plane. This latter point indicates any in-plane anisotropy to be exceedingly weak. This partial order lies in stark contrast with $B$20 helimagnets, where the cubic anisotropy dictates rigid directions of magnetic propagation at all $T$s in zero field~\cite{Muhlbauer2009,Grigoriev2005,Grigoriev2006,Bauer2017}. The order seen in GVS is more reminiscent of liquid crystals, but instead of fluctuations of molecular orientations in the real space, here the orientational disorder is reflected by a broad $q$ distribution.

To study the nature of the spiral order in detail, polarized SANS experiments were done at $T$=11~K. A small, $B$=10\,mT magnetic guide field was applied along [111] to maintain a longitudinal neutron spin polarization axis parallel to the beam, and which leaves intact the cycloidal phase in all domains \cite{Kezsmarki2015}. The relevant terms contributing to the scattered intensities in the non-spin-flip (NSF) and spin-flip (SF) processes are \cite{Moon1969}:
\begin{align}
&I^{\pm\pm}_{NSF}\propto|b(\textbf{q})\pm m_z^\perp(\textbf{q})|^2, \nonumber\\
&I^{\pm\mp}_{SF}\propto|m_x^\perp(\textbf{q})\pm im_y^\perp(\textbf{q})|^2,
\end{align}
where $m_x^\perp(\textbf{q})$, $m_y^\perp(\textbf{q})$ and $m_z^\perp(\textbf{q})$ are the $x$, $y$ and $z$ components of the Fourier transform of the local magnetization normal to $\textbf{q}$, and $b(\textbf{q})$ is the Fourier transform of the nuclear scattering length. Here we define the Cartesian $z$-axis as parallel to the neutron spin polarization axis ($\textbf{P}_{z}$). According to these equations, the NSF intensity is due to both nuclear scattering, and magnetic scattering from spin components $\parallel z$, and simultaneously normal to $\textbf{q}$. In contrast, the SF intensity can be assigned exclusively to magnetic scattering from spin components normal to both $z$ and $\textbf{q}$.

\begin{figure}[t]
\includegraphics[width=0.46\textwidth]{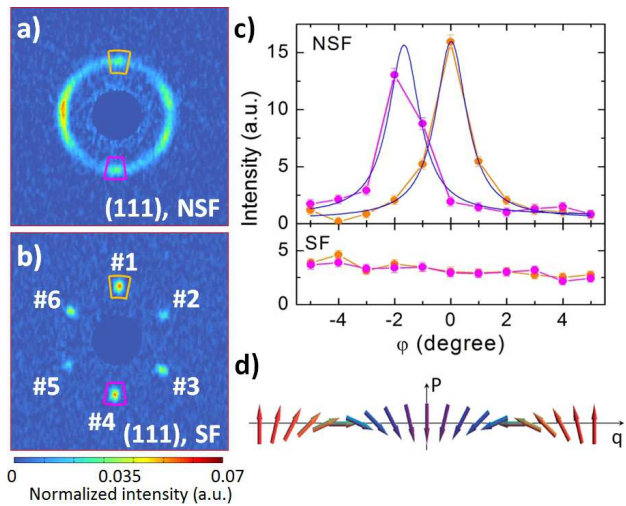}
\caption{(color online) a) Non-spin-flip (NSF) and b) spin-flip (SF) SANS scattering patterns measured on the (111) plane at 11\,K in 10\,mT. c) Rocking curves for the orange and magenta boxes shown in panel a) and b) for a sample rotation around the horizontal, [11$\overline{2}$] axis. d) Schematic of a regular cycloidal magnetic structure with spins rotating in a plane defined by $\textbf{q}$ and the rhombohedral axis $\textbf{P}$.}
\label{Fig2}
\end{figure}

Figures~\ref{Fig2}(a) and (b) respectively show the NSF and SF scattering patterns observed in the (111) plane. The ring of magnetic scattering, which originates from the [111] structural domain with rhombohedral axis $\parallel z$, contributes exclusively to NSF scattering, while the scattering from the other three structural domains contributes to both NSF and SF scattering. Since the ring of scattering appears only in the NSF channel, this confirms its origin due to spin components parallel to both $z$ and hence rhombohedral axis of the [111] structural domain. The absence of SF scattering from this structural domain rules out further spin components normal to both $\textbf{q}$ and $z$, which would be finite for helical order. Instead the data are consistent with scattering from a cycloid magnetic order, where moments rotate within a plane defined by $\textbf{q}$ and the rhombohedral axis [Fig.~2(d)].


Fig.~\ref{Fig2}(c) shows polarized SANS rocking curves for sample rotations around the horizontal [11$\overline{2}$] axis, and for intensity regions defined by the orange and magenta boxes in Figs.~\ref{Fig2}(a) and (b). These data show clearly that the ring of scattering in the (111) plane is solely responsible for the peak in the NSF intensity. From Lorentzian fits of the peaks, the correlation length along the rhombohedral axis is estimated as $\xi_\perp=2/(\text{FWHM}\cdot|q|)\sim$170\,nm, where $\text{FWHM}$ denotes the full width at half maximum of the peak, and $|q|$ the length of the scattering vector \cite{Note1}. The value of $\xi_\perp$ is comparable with that reported for high purity MnSi samples \cite{Muhlbauer2009}, indicating well developed magnetic order along the rhombohedral axis.

From the other type of structural domains with polar axes spanning 71$^{\circ}$ with $\textbf{P}_{z}$, the rocking curves trace the intensity ring out of the detector plane, and so the magnetic scattering is always finite, and almost rocking angle-independent. In the NSF rocking curve, it forms a baseline underneath the peak, while in the SF rocking curve, only the baseline is observed. The relative ratio of the baseline NSF:SF scattering is estimated to be $\sim$0.17, which is in relatively good agreement with the ratio of 1:8 expected in accord with a regular cycloid model [Fig.~2(d)]. The cycloidal nature of the spin pattern is also supported by our spin polarized experiments on the (110) plane (see the supplement \cite{Supplement}). Since the cycloidal order emerges as a consequence of the pattern of the DMI couplings specific for the C$_{3v}$ polar crystal symmetry of GVS \cite{Kezsmarki2015}, the result strongly suggests that the SkL observed in finite field can only be of the N\'eel-type and not a Bloch-type SkL. Indeed, the same distinction was found between the SF and NSF channel in the SkL phase of GVS (see the supplement \cite{Supplement}), which also confirms the SkL state is formed by cycloids in GVS.

\begin{figure}[t]
\includegraphics[width=0.46\textwidth]{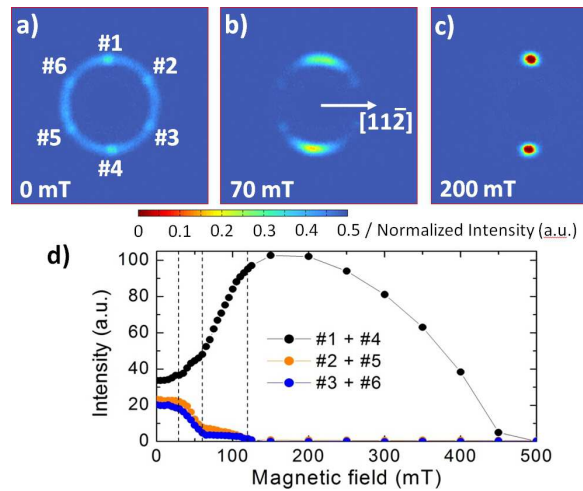}
\caption{(color online) SANS patterns measured at 11\,K in the (111) plane at a) 0\,mT, b) 70\,mT, and c) 200\,mT, when $\textbf{B}||$[11$\overline{2}$]. d) $B$-dependence of the SANS intensity for the spot pairs defined in panel a). Dashed lines indicate phase boundaries discussed in Ref.~\cite{Kezsmarki2015}}
\label{Fig3}
\end{figure}

For further evidence in support of the cycloid model, in Figs.~\ref{Fig3}(a) to (c) we show unpolarised SANS data obtained from the (111) plane when a transverse $B\parallel[11\overline{2}]$ was applied normal to the rhombohedral axis of the [111] domain. After zero-field cooling [Fig.~\ref{Fig3}(a)], the ring of scattering is observed along with the spot pairs \#1-\#4, \#2-\#5 and \#3-\#6 allocated to the remaining domains with rhombohedral axes along [$\overline{1}$$\overline{1}$1], [$\overline{1}$1$\overline{1}$] and [1$\overline{1}$$\overline{1}$], which respectively span 19.5$^{\circ}$, 61.9$^{\circ}$ and 61.9$^{\circ}$ with $B$. In these latter three domains, the component of $B$ along the rhombohedral axis drives the cycloidal-SkL and SkL-FM phase transitions (indicated by dashed lines in Fig.~\ref{Fig3}(d)) as discussed in Ref.~\cite{Kezsmarki2015}. In the [111] domain however, the intensity distributed around the ring is gradually collected into a spot pair with $\textbf{q}$ normal to $B$ [Figs.~\ref{Fig3}(b) and (c)], and this modulated state survives up to high field [Fig.~\ref{Fig3}(d)]. This provides further evidence that cycloidal order is realised in GVS, since the susceptibility of a cycloid is generally enhanced normal to the cycloid plane containing $\textbf{q}$. Under an in-plane $B$, the cycloid order transforms smoothly into a transverse conical order with $\textbf{q}$ normal to $B$. Here the cone axis is parallel to $B$, and the cone angle closes at $\sim$0.45\,T [Fig.~\ref{Fig3}(d)]. In contrast with cubic helimagnets where $B$, $\textbf{q}$ and the cone axis are all parallel with each other and this longitudinal conical phase covers the major part of the phase diagram for any $B$ direction \cite{Muhlbauer2009,Grigoriev2006,Bauer2017}, in GVS the transverse ($\textbf{B}\perp\textbf{q}$) conical state exists only when $B$ lies close to the rhombohedral plane within which $\textbf{q}$ is confined.

\begin{figure}[t]
\includegraphics[width=0.46\textwidth]{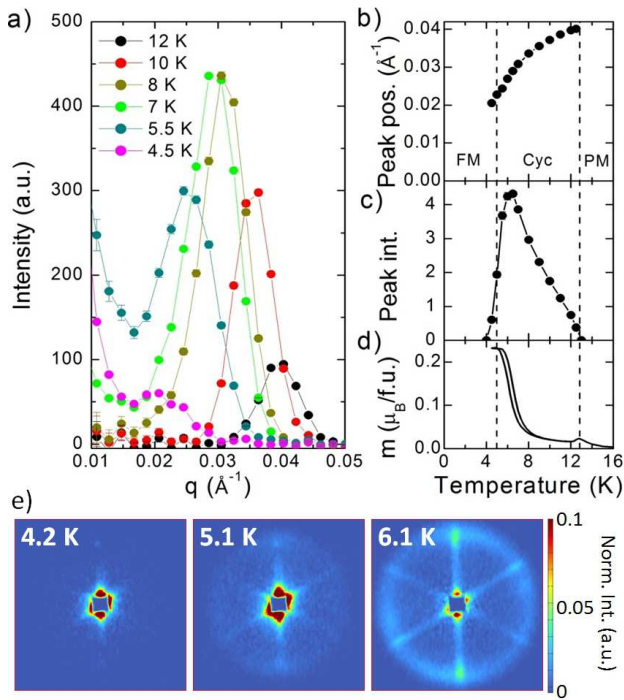}
\caption{(color online) a) $T$-dependence of the $|q|$-dependent, azimuthally-averaged SANS intensity on the detector plane. b) $T$-dependence of the cycloid scattering vector and c) its intensity. d) $T$-dependence of the magnetization measured in 10\,mT. e) Typical unpolarised SANS patterns measured in the vicinity of the ferromagnetic-cycloidal transition.}
\label{Fig4}
\end{figure}

Finally, Fig.~\ref{Fig4} shows unpolarised SANS data summarizing the $T$-evolution of the cycloidal order at $B$=0\,T. Upon cooling below $T_C$=13\,K the peak intensity at finite $|q|$ increases smoothly [Fig.~\ref{Fig4}(c)] implying a continuous transition from the paramagnetic to cycloidal state. As shown in Figs.~\ref{Fig4}(a) and (b) this peak is initially located at $|q|$=0.04\,{\AA}$^{-1}$, with its position $|q|$ falling by 50\% on cooling. In the low $T$ range where $|q|$ is still finite, the SANS intensity drops sharply to zero [Fig.~\ref{Fig4}(c)] and simultaneously a tail of low-q scattering develops indicating growing FM correlations [Fig.~\ref{Fig4}(a)]. This phase coexistence, sharp change in SANS intensity, and hysteresis in the $T$-dependence of the magnetization [Fig.~\ref{Fig4}(d)], indicate a first-order character of the cycloidal-FM transition consistent with recent specific heat data~\cite{Widmann2016}. On the other hand, as shown in Fig.~\ref{Fig4}(a) when close to $T_S$ the radial distribution of the intensity significantly broadens in $|q|$, and the intensity ring transforms into a disk-like profile in the (111) plane with a sharp cut off at the high-q end [Fig.~\ref{Fig4}(e)]. In the other three domains, the intensity disks appear as a six-spoked asterix when imaged by SANS along the [111] axis. These data show that close to the cycloidal-FM transition the length of $|q|$ varies strongly, or even fluctuates, in the rhombohedral plane. This could indicate that the transition may be continuous in theory, but hysteresis effects emerge due to slow dynamics when the rearrangements of large structures are involved \cite{Butykai2017tobe}, as also found recently in MnSi \cite{Bauer2017}.

Nevertheless, in comparison with the general phase diagram of the $B$20 helimagnets, the results reported here for GVS stand out as rather unusual. At such low $T$s in GVS, it is unlikely that the microscopic couplings such as the symmetric exchange or DMIs vary sufficiently to drive the cycloidal-FM transition at $T_{S}$. A more plausible scenario is that thermal fluctuations are crucial for stabilizing the long-wavelength modulated states close to $T_{C}$, and when they are suppressed at low $T$ the system transforms to a FM state expected in accord with the strong easy-axis magnetic anisotropy~\cite{Kezsmarki2015}. Similar behaviour has been discussed in the context of the spiral-FM transition in the elemental rare-earths like Dy and Tb, wherein the effective anisotropy becomes more influential as fluctuations are suppressed at low $T$ \cite{Izyumov1984}. In those systems the strong axial anisotropy leads to a larger periodicity in real-space (shorter $|q|$) with increased anharmonicity, and close to transition the spiral transforms to a soliton lattice with periodically arranged FM domain walls. Such a state in GVS may explain the unusual $T$-dependent SANS data shown in Fig.~\ref{Fig4}(e), and indeed in Fig.~\ref{Fig4}(b) we find that $|q|$ reduces by 50\% on cooling from $T_{C}$ to $T_{S}$. This proposed $T$-dependent interplay between the easy-axis anisotropy and thermal fluctuations likely governs the fine structure of the phase diagram in GVS, again in contrast with the $B$20s which have a comparatively weaker anisotropy, and in which $|q|$ changes at most 10\% upon cooling towards a low $T$ ground state that always remains helically modulated \cite{Grigoriev2006}.

In conclusion, we have evidenced GaV$_4$S$_8$ (GVS) to display cycloidally-modulated magnetic order at zero magnetic field ($B$), consistent with the formation of the reported N\'eel-type skyrmion state under finite $B$. As $T\rightarrow0$, cycloidal order collapses and a FM ground state is realized. This suggests thermal fluctuations to be crucial for stabilizing the modulated states in GVS, while FM order emerge at low $T$ due to the strong easy-axis magnetic anisotropy. At the same time however, in finite $B$ this anisotropy suppresses the competing conical phase, promoting an extended thermal stability range of the N\'eel-type skyrmion state in GVS compared with the tiny stability range of the skyrmion state in the $B$20s.




We wish to thank N.~Reynolds, L.~R\'ozsa, U.K.~R\"ossler, M.~Mark\'o, and in particular I.~K\'ezsm\'arki, A.~Loidl, H.M.~R\o{}nnow for fruitful discussions and support. This work was supported by Hungarian Research Funds OTKA K 108918, OTKA PD 111756, Bolyai 00565/14/11, by the Lend\"ulet Program of the Hungarian Academy of Sciences, the Swiss National Science Foundation Sinergia network NanoSkyrmionics, and the Swiss National Science Foundation (SNF) project grant 153451. This work is based on neutron experiments performed at the Swiss spallation neutron source SINQ, Paul Scherrer Institute (PSI), Villigen, Switzerland, and the Institut Laue-Langevin (ILL), Grenoble, France.


\end{document}